\documentclass[12pt]{article}
\usepackage{epsf}

\def\1ad{\mbox{\normalsize $^1$}}
\def\2ad{\mbox{\normalsize $^2$}}
\def\3ad{\mbox{\normalsize $^3$}}
\def\4ad{\mbox{\normalsize $^4$}}
\def\5ad{\mbox{\normalsize $^5$}}
\def\6ad{\mbox{\normalsize $^6$}}
\def\7ad{\mbox{\normalsize $^7$}}
\def\8ad{\mbox{\normalsize $^8$}}

\def\npb#1#2#3{{\it Nucl. Phys.} {\bf B#1} (#2) #3 }
\def\plb#1#2#3{{\it Phys. Lett.} {\bf B#1} (#2) #3 }
\def\prd#1#2#3{{\it Phys. Rev. } {\bf D#1} (#2) #3 }
\def\prl#1#2#3{{\it Phys. Rev. Lett.} {\bf #1} (#2) #3 }

\def\cmp#1#2#3{{\it Commun. Math. Phys.} {\bf #1} (#2) #3 }

\def\bb#1{{\tt hep-th/#1}}

\def\jhep#1#2#3{{\it J. High Energy Phys.} {\bf #1} (#2) #3 } 
\def\atmp#1#2#3{{\it Adv. Theor. Math. Phys.} {\bf #1} (#2) #3 }
\def\bb#1{{\tt hep-th/#1}}

\def\CL{{\cal L}}   
  \def\CD{{\cal D}} 
 
\def\CN{{\cal N}}

\def\dj{\hbox{d\kern-0.347em \vrule width 0.3em height 1.252ex depth
-1.21ex \kern 0.051em}}

\def\half{{1\over 2}\,}

\def\pt{\partial}

\def\ord{{\cal O}}

\newcommand{\be}{\begin{equation}}
\newcommand{\ee}{\end{equation}}
\newcommand{\ba}{\begin{eqnarray}}
\newcommand{\ea}{\end{eqnarray}}
\newcommand{\nn}{\nonumber}

\newcommand{\ie}{\mbox{{\em i.e.~}}}

\def\half{\frac{1}{2}}

\def\shalf{{\mbox{$\half$}}}

\newcommand{\ben}{\begin{equation*}}
\newcommand{\een}{\end{equation*}}
\newcommand{\ban}{\begin{eqnarray*}}
\newcommand{\ean}{\end{eqnarray*}}
\newcommand{\brr}{\begin{array}}
\newcommand{\err}{\end{array}}
\newcommand{\bc}{\begin{center}}
\newcommand{\ec}{\end{center}}

\begin{document}

\newcommand{\sheptitle}
{Aspects of Instanton Dynamics in AdS/CFT  Duality}

\newcommand{\shepauthor}
{
J.L.F.~Barb\'on$^{a,}$\footnote[1]{J.Barbon@phys.uu.nl}, 
A.~Pasquinucci$^{b,}$\footnote[2]{Andrea.Pasquinucci@cern.ch}}

\newcommand{\shepaddress}
{
$^{a}$ Spinoza Institute, Leuvenlaan 4, 3584 CE, Utrecht, The Netherlands\\
$^{b}$ Physics Department, Milano University, via Celoria 16, I-20133
Milano, Italy}

\newcommand{\shepabstract}
{We consider aspects of  instanton dynamics in the large-$N$ limit 
using the AdS/CFT duality for D0/D4 bound states. 
In the supergravity picture of wrapped D0-brane world-lines on
D4-branes, we find the single-instanton measure and discuss 
its dependence on compactification finite-size effects, as well as 
its matching to perturbative results. 
In the non-supersymmetric case, the same dynamical effects that produce 
the theta-angle dependence perturbatively in $1/N$, render the 
instantons unstable, although approximate instantons of very small 
size still exist. 

The smeared D0/D4 black-brane supergravity solution can be interpreted 
as dual to a field theory configuration of an instanton condensate in the 
vacuum. In this case, we derive a holographic relation between 
the bare theta angle and the topological charge density of the instanton 
condensate. 
}

\begin{titlepage}
\begin{flushright}
IFUM-642/FT\\
Bicocca-FT-99-08\\
SPIN-99-10\\
hep-th/9904190\\

\end{flushright}
\vspace{1in}
\begin{center}
{\large{\bf \sheptitle}}
\bigskip\bigskip \\ \shepauthor \\ \mbox{} \\ {\it \shepaddress} \\ 
\vspace{1in}
{\bf Abstract} \bigskip \end{center} \setcounter{page}{0}
\shepabstract
\vspace{1in}
\begin{flushleft}
April 1999
\end{flushleft}


\end{titlepage}

\newpage

%
%

\section{Introduction}
\setcounter{equation}{0}

As approximation schemes for gauge dynamics,
 instanton calculus \cite{shif}  
and 't Hooft's $1/N$ expansion \cite{tof} do not seem to combine
in a useful fashion.  
 Since effects of a charge $k$ instanton
sector are of $\ord (e^{-8\pi^2 k/g^2}) = \ord (e^{-N})$, it would seem
that they are always irrelevant in the large-$N$ limit
unless they control the {\it leading} contribution to some observable
(for instance, because of supersymmetry non-renormalization arguments),
or somehow the integral over instanton moduli space is ill-defined.   
Such non-commutativity of the large-$N$ limit and the instanton sum
is assumed to be behind well-known instances of theta-angle dependence
at perturbative order  in the $1/N$ expansion, notably in the
context of large-$N$ chiral dynamics \cite{venewitt}.  

On the other hand,
it is known that some toy models \cite{toym} completely
suppress instanton-like excitations once the large-$N$ limit has been
taken. In other words, the `effective action' resulting after large-$N$
diagram summation does not support instantons any more. So, one may
wonder whether the large-$N$ `master field'  always loses its
discrete topological structure. Recently, the AdS/CFT correspondence of
\cite{Malda,gkpw} has provided a new set of non-trivial master fields
for some gauge theories. In particular, the theta-angle dependence
of $\CN=4$ Super Yang--Mills $({\rm SYM}_4)$ in four dimensions 
can be studied in the large-$N$ expansion via perturbative Type IIB string
theory in ${\bf AdS}_5 \times {\bf S}^5$. It is saturated by instantons,
which appear in the supergravity description as D-instantons
\cite{dinst}. So, the radical view that an instanton gas is incompatible
with the large-$N$ limit is not vindicated in this case.

In this paper, 
we investigate these questions in a QCD cousin introduced by Witten
\cite{wit2}, which admits a supergravity description of its master
field, while removing the constraints of extended supersymmetry and
conformal invariance.   
More precisely, we would like to learn under what conditions some
kind of topological configurations  (instantons) 
still give the leading semiclassical effects of  $\ord (e^{-N})$,
even after the planar diagrams have been summed over. We shall 
focus on the most elementary case of the dilute limit, \ie the 
single-instanton sector. 

One description of this theory is in terms of $N$ D4-branes wrapped on 
a circle ${\bf S}^1_\beta$ of length $\beta$,
with thermal boundary conditions. At weak coupling, the low-energy
theory on the D4-branes is a perturbative five-dimensional Super
Yang--Mills theory $({\rm SYM}_{4+1})$,
which reduces to  non-supersymmetric, $SU(N)$ Yang--Mills theory
in four euclidean dimensions $({\rm YM}_4)$,
at distance scales much larger than the inverse temperature $T^{-1}=
\beta$.
Since five-dimensional gauge fields originate from massless open
strings, their coupling scales as
$g_5^2 \sim g_s\,\sqrt{\alpha'}$ with 
$g_s = {\rm exp}(\phi_{\infty})$ denoting the
string coupling constant. Therefore, the four-dimensional coupling at
the cut-off scale $T$ is given by $g^2 \sim g_s \,T\,\sqrt{\alpha'}$.

The weak-coupling description of instantons in this set up is in terms
of  D0/D4-branes bound states. Due to the Wess--Zumino coupling 
between the type IIA Ramond--Ramond (RR) one-form
 and the gauge fields on the
D4-branes world-volume, $ \CL_{\rm WZ} =   
C_{\rm D0} \wedge F\wedge F$,  
a D0-brane `inside' a D4-brane
carries the instanton charge. The action of an euclidean world-line
wrapped on a circle ${\bf S}_{\beta}^1$ is
\be\label{d0oldaction}
S_{{\rm D}0} = M_{{\rm D}0} \cdot \beta = {\beta \over \sqrt{\alpha'}
\,g_s}=
{8\pi^2 \over g^2} \equiv {8\pi^2 N \over \lambda} 
.\ee
Incidentally, this relation also fixes the numerical conventions in the
definition of the four-dimensional coupling $g$. 
We have also introduced the standard notation for the large-$N$ 't Hooft
coupling $\lambda \equiv g^2 N$. 

The moduli of these instantons are encoded in the 
quantum mechanical zero-modes  of the D0--D0
 and D0--D4 strings. For a standard
compactification, the D0-branes (\ie the 
`instanton particles' of the gauge
theory) describe standard  instantons of $\CN=4$ ${\rm
SYM}_4$ (see \cite{usinst} for some generalizations).   
If the circle breaks supersymmetry, the instanton fermionic
 zero modes should be lifted
accordingly to mass of $\ord(T)$, and one should get essentially a
Yang--Mills instanton with no fermionic zero modes. 
Other one-loop effects would incorporate the perturbative running of
the coupling constant in the standard way.

The supergravity framework for ${\rm SYM}_{4+1}$ at finite temperature 
is given by the   black D4-brane solution \cite{wit2, mali}.
  The full metric in the string
frame is:\footnote{See, for example, \cite{krev} and references therein
 for a review of metrics relevant to this paper.}   
\be
\label{fulmet}
ds^2 = H_4^{-\half} (h\,d\tau^2 + d{\vec y}^{\,2}) + H_4^{\half}
\,\left( dr^2 / h + r^2 \,d\Omega_{4}^2 \right)
\ee
where
\be
\label{prof}
H_4= 1+ (r_{Q4}/r)^3 \,,\;\;\;\;\;\;\; h=1-(r_0 /r)^3.
\ee
There are two length scales associated with this metric: the
Schwarzschild radius, $r_0$,  related to the Hawking temperature $T$ by
$ T^{-1} = \beta = (4\pi/3)\, r_0 \,[H_4 (r_0)]^{1/2}$,
and the charge radius $r_{Q4}$, given by
\be
\label{crad}
r_{Q4}^3 = - \shalf r_0^3 + \sqrt{{\mbox{$1\over 4$}} r_0^6 + 
\alpha'^{\,3}\,(\pi\, g_s\,N)^2},
\ee

In the Maldacena or gauge-theory limit, one scales $\alpha' \rightarrow 0$ 
with $r/\alpha' =u $ and $r_0 /\alpha' = u_0$ 
fixed.  The new coordinate $u$ has dimensions of energy and the scaling
properties of the Higgs expectation value. In this limit, only the
combination         
\be
\label{nea}
\alpha'^{\,2} \,H_4 \rightarrow {\pi \,g_s \,N\,\sqrt{\alpha'}
\over u^3} = {\lambda \,\beta\over 8\pi\, u^3}
\ee
is relevant. In the supergravity picture, the D4-branes have disappeared
in favour of the `throat geometry' ${\bf X}_{\rm bb}$
 (\ref{fulmet}), \ie we have no open
strings and the description is fully gauge invariant. The black-brane  
manifold ${\bf X}_{\rm bb}$, with topology 
${\bf R}^2 \times {\bf R}^{4} \times {\bf S}^4$,
has a boundary at $u=\infty$ of topology ${\bf S}^1 \times {\bf R}^4$,
which is interpreted as the ${\rm SYM}_{5}$ 
space-time (the $(\tau, {\vec y})$ space). The physical
interpretation is that asymptotic boundary conditions for the
supergravity fields at $u=\infty$ represent coupling constants of
microscopic operators in  the   gauge theory \cite{gkpw}.    

The same boundary conditions are satisfied by the extremal D4-brane
metric with thermal boundary conditions. This is 
 the `vacuum' manifold, denoted ${\bf X}_{\rm vac}$, with
topology ${\bf S}^1 \times {\bf R}^5 \times {\bf S}^4$, obtained from
(\ref{fulmet}) by
setting $u_0 =0$, with {\it fixed} $\beta$. However,
one can show  \cite{wit2,bkr} that  ${\bf X}_{\rm vac}$ is suppressed by a
relative factor of $\ord (e^{-N^2})$ with respect to ${\bf X}_{\rm bb}$,
 in the large-$N$ limit. In other words, the $\ord (N^2)$ actions
satisfy
\be
\label{balan}
I({\bf X}_{\rm bb}) - I({\bf X}_{\rm vac}) = -K\,N^2 \,\lambda \, VT^4 <0
\ee
for any $T>0$, with $K$ a positive constant, \ie
 there is no Hawking-Page 
transition \cite{hp,wit2}.  

Unlike the case of $\CN =4$ ${\rm SYM}_4$, the dilaton is not constant
in the supergravity description. It becomes strongly coupled at radial
coordinates of order  
$u\sim \ord (N^{4/3} /\beta\,\lambda)$, where one has $e^{\phi} = g_s
\,(H_4)^{-1/4} =\ord (1)$. Beyond this point, one should
use a dual picture in terms of a wrapped M5-brane in M-theory, \ie a
quotient of ${\bf AdS}_7 \times {\bf S}^4$. For the
purposes of the discussions in
 this paper, we are studying the theory at fixed energy scales of $\ord
(1)$ in  the 't Hooft's
large-$N$ limit, with fixed $\lambda=g^2 N$. Therefore, such non-perturbative 
thresholds effectively decouple in the regime of interest,   
and we shall  formally extend
the D4-brane manifold all the way up to $u=\infty$.

{}From a physical point of view, $\alpha'$-corrections to the classical
geometry pose a more serious limitation to the supergravity description.
The curvature at the horizon scales as 
$( u_0 \,g_s\,N
\,\sqrt{\alpha'})^{-1/2} \sim \lambda^{-1}$, 
in string units, 
so that the supergravity description is accurate only for large bare
't Hooft coupling $\lambda \gg 1$. On the other hand, the glueball mass
gap \cite{oo} in this theory is of order
$
M_{\rm glue} \sim \beta^{-1}
=T$, 
while inspection of the Wilson loop expectation value gives a
four-dimensional string tension \cite{stten} of order 
$\sigma \sim \lambda \,T^2$, \ie hierarchically larger in the
supergravity regime. This lack of scaling 
indicates that the supergravity picture is far from the `continuum limit'
of the ${\rm YM}_4$ theory, a suspicion already clear from the existence
of non-QCD states of Kaluza--Klein origin at the same mass scale as the
glueballs: $M_{\rm KK} \sim T \sim M_{\rm glue}$.

\section{The Localized Instanton}
\setcounter{equation}{0}

The natural candidate for an instanton excitation in the large-$N$ 
supergravity
picture  is a D0-brane probe wrapped around
the thermal circle.  For the supersymmetric case, this is indeed the
T-dual configuration to the D-instantons in ${\bf AdS}_5 \times {\bf
S}^5$ discussed in \cite{dinst}.
Wrapped D0-branes have the correct quantum numbers to be
interpreted as Yang--Mills instantons in the effective four-dimensional
theory. The topological charge is interpreted as the wrapping number on
the thermal circle ${\bf S}^1_{\beta}$. In the large-$N$ limit, it is
justified to take the D0-brane as a probe, neglecting its back-reaction
on the supergravity fields, since the gravitational radius is 
sub-stringy: $(r_{\rm probe})^7 \sim \alpha'^{\,7/2}\,e^{\phi} \sim \ord
(1/N)$,
although for instanton numbers of $\ord (N)$
with identical moduli we may need a
supergravity description for the instanton dynamics in terms of the
D0-branes near-horizon geometry (\ie a T-dual of the limit in 
\cite{koo}, or the solution of section 3 below). We shall postpone
 these
interesting 
complications by working in the single-instanton sector, and with
instanton moduli of $\ord (1)$ in the 't Hooft large-$N$ limit.

One important ingredient of the 
 the instanton/D0-brane mapping  
is a physical interpretation in gauge-theory language of the 
 wrapped D0 world-line's radial position.
 For this purpose, we use the generalized
UV/IR connection as discussed in \cite{uvir}. According to this, a
radial coordinate $u$ is associated to a length scale 
 in the ${\rm SYM}_{p+1}$ gauge theory
of order $ \ell\sim \sqrt{ g_{p+1}^2 \,N /u^{5-p}}$. 
Thus, in our case, the size parameter $\rho$ of the instanton satisfies:
\be
\label{sizep}
\rho^2 = {\beta\,\lambda\over u}
.\ee
We will assume this relation as the definition of the
instanton's size modulus. 

We will now discuss both manifolds with ${\bf S}^1 \times {\bf R}^4$
boundary conditions at $u=\infty$, in spite of the fact that eq.\ 
(\ref{balan}) ensures the dynamical dominance of ${\bf X}_{\rm bb}$. 
The reason for considering also the `vacuum' manifold is first that we 
find interesting differences between the manifolds, and that 
${\bf X}_{\rm vac}$ is the only relevant manifold for supersymmetric
compactification, with which we can make contact with the 
${\bf AdS}_5 \times {\bf S}^5$ case. 

\subsection{Vacuum Manifold}

The ${\bf S}^1$ factor on the 
boundary  extends to the bulk of ${\bf X}_{\rm vac}$ becoming
singular as $u\rightarrow 0$, since   
${\rm Vol}({\bf S}^1_u) = \beta \sqrt{g_{\tau\tau}} = \beta\,(H_4)^{-1/4}   
\rightarrow 0$.  
However, the action of the instanton is constant, due to the dilaton
dependence in the Dirac--Born--Infeld action:
\be
\label{dbis}
S_{\rm D0} = M_{\rm D0} \int_{0}^{\beta} d\tau\, (g_s\,e^{-\phi})
\,\sqrt{g_{\tau\tau}}
= {8\pi^2 \over g^2}
.\ee
Thus, the size $\rho$ is an exact modulus in the supergravity
description on ${\bf X}_{\rm vac}$. 
On general grounds, the path integral of a D-particle in a curved
background ${\bf X}$ 
contains an ultralocal term in the measure of the form $\CD X^{\mu}
\,[{\rm det}(g_{\mu\nu})]^{1/2}$, to ensure invariance under target-space
diffeomorphisms. In the description of instantons on the manifold ${\bf
X}$, we concentrate on the
zero-mode part which then leads to a single-instanton measure
\be
\label{measurex}
d\mu \,({\bf X}) = C_{N,\lambda} \;d\eta\;\int_{{\bf S}^1_{\beta} \times {\bf
S}^4_{\Omega} } (\alpha')^{-5}\;d{\rm Vol}({\bf X}),
\ee
where $d\eta$ is the measure over fermionic zero-modes (sixteen in the
supersymmetric case), and $C_{N,\lambda}$ is
a constant to be determined by the matching to the perturbative measure.
We have produced a measure in the physical space-time and
scale-parameter space by averaging 
 over ${\bf S}^1_{\beta} \times {\bf S}^4_{\Omega}$.  
The result for ${\bf X}_{\rm D4} = {\bf X}_{\rm vac}$,
 using the UV/IR connection (\ref{sizep}) is:
\be
\label{measud4}
d\mu \,({\bf X}_{\rm D4}) = C_{N,\lambda} \; \lambda^5 \, 
(\rho \,T)^{-6} \; \rho^{-5}\,d\rho\,d{\vec y} \,d\eta.
\ee
We  see that the presence of the dimensionful scale $T$
explicitly violates the conformal invariance of the measure, which we
must take as a concrete prediction of the supergravity approach. As
such, it is valid at large $N$ and $\lambda$.

The singularity of ${\bf X}_{\rm vac}$ as $u\rightarrow 0$ 
is not relevant. At $u\sim u_s =
T \lambda^{1/3}$ the size of the world-line is of $\ord (1)$ in string
units. So, for $u\ll u_s$ we must use the T-dual metric of $N$ D3-branes
smeared over the dual circle of coordinate
length ${\widetilde \beta} = 4\pi^2 \alpha' /\beta$:
\be
\label{tdual}
ds^2 ({\bf X}_{\widetilde {\rm D3}}) = H_4^{-\half} \,
 d{\vec y}^{\,2} + 
 H_4^{\half}\, \left( d{\widetilde \tau}^2 + dr^2 + r^2
\,d\Omega_4^2 \right)
,\ee
with ${\widetilde \tau} \equiv {\widetilde \tau} + {\widetilde 
\beta}$.
In the T-dual metric,\footnote{Notice that the UV/IR connection  
(\ref{sizep}) remains unchanged by T-duality, as the new metric
only differs by $g_{\tau\tau} \rightarrow 1/g_{\tau\tau}$, with
the $u,{\vec y}$ components  of the metric unaffected.}
the size of ${\bf S}^1_u$ grows with decreasing $u$.  

In fact, the metric (\ref{tdual}) is unstable if any small amount
of energy is added. It collapses to the array solution of localized
D3-branes \cite{greglaf}: 
$$
ds^2 ({\bf X}_{\rm D3}) = H_3^{-\half} d{\vec y}^{\,2} + H_3^{\half}\,
\left(d{\widetilde\tau}^2 + dr^2 + r^2 \,d\Omega_4^2 \right),
\;\;\;\;{\rm with} \;\;\;
H_3 = 1+\sum_n {4\pi\,{\widetilde g}_s \,\alpha'^{\,2} 
\over |r^2 + (n{\widetilde\beta})^2 |^2}.
$$
By the T-duality rules and our coupling conventions: $4\pi{\widetilde g}_s =
8\pi^2 \,g_s \sqrt{\alpha'}/\beta = g^2$. 
In the regime $r\gg {\widetilde \beta}$ we
can approximate the discrete sum over images by a continuous integral,
and
we recover the smeared metric (\ref{tdual}) as an approximation. On the
other  hand, for $r\ll {\widetilde \beta}$ we can instead neglect the
images and approximate the sum by the $n=0$ term. The result is of
course the standard ${\bf AdS}_5 \times {\bf S}^5$ metric corresponding
to D3-branes at strong coupling. Indeed, the UV/IR relation for
D-instantons in D3-branes \cite{dinst}, 
$ 
\rho = \sqrt{\lambda} / u
,$ 
matches the five-dimensional one (\ref{sizep}) precisely at $u=u_{\rm
loc} = 1/\beta$, which is equivalent to $r=r_{\rm loc} =
{\widetilde \beta}/4\pi^2$.     

 The instanton measure (\ref{measud4}) matches across these
finite-size transitions to the corresponding measures for the new  
manifolds ${\bf X}_{\widetilde {\rm D3}}$ and ${\bf X}_{\rm D3}$,
because the  definition (\ref{measurex}) applies in general and the
volume form matches across the transitions at $u=u_s$ and $u=u_{\rm
loc}$. The resulting measures are (both up to $\ord (1)$ numerical
 factors): 
\ba
\label{meas}
d\mu \,({\bf X}_{\widetilde {\rm D3}}) &=& C_{N,\lambda}  \;\lambda^4
\;(\rho \,T)^{-3} \;\rho^{-5}\,d\rho\,d{\vec y}\,d\eta,
\nn \\  
d\mu \,({\bf X}_{\rm D3}) &=& C_{N,\lambda} \; \lambda^{5/2} \;\rho^{-5}\,
d\rho\,
d{\vec y}\, d\eta
.
\ea
This last measure is  conformally invariant, and coincides with
that of refs. \cite{dinst} for D-instantons in ${\bf AdS}_5 \times
{\bf S}^5$.   
 
Finally, 
as pointed out in the introduction, the validity of the supergravity
picture is limited by the requirement that we can control the
$\alpha'$-corrections. The curvature of the D4-brane metrics
 is of  $\ord (1)$ in string units
 at the
`correspondence line' $u_c \sim (\beta \,\lambda)^{-1}$
\cite{horpol}. For the D3-brane metrics, the condition is simply
$\lambda 
\sim 1$. This implies that, for $\rho < \beta$, we have a
correspondence line for instanton sizes
$ 
\rho =\rho_c = \beta\,\lambda.
$
For $\rho >\beta$ the correspondence line is independent of $\rho$ and
lies at $\lambda \sim 1$. 
Below the correspondence line the system is better described in
Yang--Mills perturbation theory, although we lose the analytic control
over the $1/N$ expansion.

 The geometrical D-instanton measure in ${\bf AdS}_5 \times {\bf S}^5$  
 has been matched to the perturbative instanton measure
in the $\CN=4$ ${\rm SYM}_4$ theory in great detail, including
multi-instanton terms \cite{valya}. In particular, this allows us
to fix the coupling-dependent constant as $C_{N,\lambda} = N^{-7/2} \,
\lambda^{3/2}$.  This is rather remarkable, since the geometrical
measure  holds
at large $\lambda$, whereas the perturbative measure is derived in
Yang--Mills perturbation theory, valid for $\lambda \ll 1$. This robustness
of the instanton measure in this case might be due to the high
degree of supersymmetry and/or conformal symmetry. For instance,
the analogous matching between the D4-brane supergravity measure
(\ref{measud4})  
and the perturbative description of the 
 `instanton particles' of ${\rm SYM}_5$, through the
correspondence line $\rho = \rho_c = \beta \,\lambda$,  fails  by one
power of $\lambda$. This means that the very precise matching of
\cite{valya} for  
 ${\bf AdS}_5$ D-instantons is probably a consequence of conformal
invariance.

This discussion may be summarized in Fig. 1, where the finite-size
transitions, as well as the correspondence lines are depicted as a
function of the 't Hooft coupling and the instanton size.  
 
\begin{figure}
\hspace*{1.4in}
\epsfxsize=3in
\epsffile{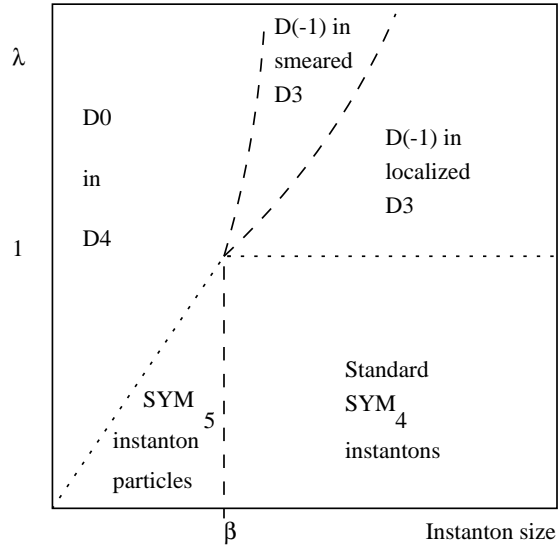}
\caption{
\small
Instanton phase diagram for the compactified D4 theory on a
supersymmetric circle of size $\beta$. 
 The dotted lines
denote the correspondence curves separating the geometric descriptions at
large 't Hooft coupling $\lambda =g^2 N$ from the perturbative  
SYM descriptions at small $\lambda$.  
 Dashed lines represent transitions driven by the finite size
of the compactification circle. }
\end{figure}

\subsection{Black-brane Manifold}

Although the wrapping charge of D0-branes is well defined in ${\bf
X}_{\rm vac}$, the thermal circle being non-contractible, this is not
the case for  ${\bf X}_{\rm bb}$, whose  
 $(\tau, u)$ subspace has ${\bf R}^2$ topology. Therefore,  the thermal
circle at fixed radial coordinate  ${\bf S}^1_u$,  is contractible, being
the boundary of a 
disc: ${\bf S}^1_u = \pt {\bf D}_u$, 
 \ie\ we can `unwrap' the
D0-brane instanton through the horizon. 
Thus, while exact
 instanton charges can be identified in the supersymmetric case, no 
quantized 
topological charge seems to survive in the non-supersymmetric case,
due to the dynamical dominace of ${\bf X}_{\rm bb}$ in the large-$N$
limit 
 (\ref{balan}). 

Still, we can talk of approximate or `constrained' instantons, provided
the probe D0-brane world-line wraps far away from the horizon. In this case
the un-wrapping costs a large action. In order to estimate the action
as a function of $u$ (or the instanton size $\rho$), we calculate the
Dirac--Born--Infeld action of the probe D0-brane:  
\be
\label{corac}
S_{\rm D0} = M_{\rm D0} \int_{0}^{\beta} d\tau\, (g_s\,e^{-\phi})
\,\sqrt{g_{\tau\tau}}
= {8\pi^2\over g^2} \sqrt{h}  = {8\pi^2 \over g^2}
\sqrt{1-(\rho/\beta)^6}
,\ee
where we have used  the UV/IR relation (\ref{sizep}) in the last step.  
Thus, $\rho$ is not an exact modulus, as instantons tend to grow.  
For an instanton of the order of the glueball's Compton
wave-length $\rho \sim \beta$, the action is comparable to the
vacuum action, and the instanton has disappeared (un-wrapped).

In the far ultraviolet, we can use the approximate instantons 
of very small size 
 $\rho \ll \beta$, 
 to measure
a  
`running effective theta angle', by requiring that the approximate
instanton is weighed by a phase ${\rm exp}(i\theta_{\rm eff})$, with
$\theta_{\rm eff} (u=\infty)=\theta$, the bare theta angle of the
four-dimensional ${\rm YM}_4$ theory.   
Following Witten \cite{thetaw}, a bare theta angle is associated to a
RR two-form 
\be
\label{witprof}
f_{\rm D0} = dC_{\rm D0} =
{\overline \theta} \,{3\over \pi \zeta^7} \,d\zeta\wedge d\psi, 
\ee
where, in the notation of \cite{thetaw},  $\zeta^2 =u/ u_0$, and
$\psi=2\pi\tau/\beta$.  
The bare theta angle, measured at $u=\infty$, 
is $\theta = {\overline \theta} \,({\rm mod} \;2\pi)$, due to the
multiplicity of meta-stable vacua as described in \cite{thetaw},
\ie $f_{\rm D0} \propto {\overline\theta} = (\theta + 2\pi n)$ 
in the $n$-th vacuum (see also \cite{ozpa} for another geometric
approach to this question). In
what follows, we shall obviate this technicality by working in the
$n=0$ vacuum, so that $\theta ={\overline \theta}$.  
The effective theta angle at throat radius $u$ is 
\be
\label{thetaf}
\theta_{\rm eff} (u) = \oint_{{\bf S}^1_u}
 C_{\rm D0} = \int_{{\bf D}_u} f_{\rm
D0}  
=\theta \left(1-6\int_{\zeta(u)}^{\infty} d\zeta
\,\zeta^{-7}\right)
= \theta \; h(u)
= \theta \left(1-(\rho/\beta)^6 \right)
.\ee

The `correspondence line' $u_c \sim (\beta \,\lambda)^{-1}$ 
\cite{horpol}, controlling $\alpha'$-corrections is also defined
in ${\bf X}_{\rm bb}$. In terms of instanton sizes,
 for $\rho < \beta$, we have a 
correspondence line at  $ \rho_c = \beta\,\lambda$.
Since no instantons survive for $\rho > \beta$ in the supergravity
picture, the finite-size effects related to T-duality in ${\bf
S}^1_{\beta}$ and localization effects are absent for ${\bf X}_{\rm
bb}$, \ie there is no phase of D-instantons in ${\bf AdS}_5 \times {\bf
S}^5$. The situation can be summarized by Fig. 2. 

\begin{figure}
\hspace*{1.4in}
\epsfxsize=3in
\epsffile{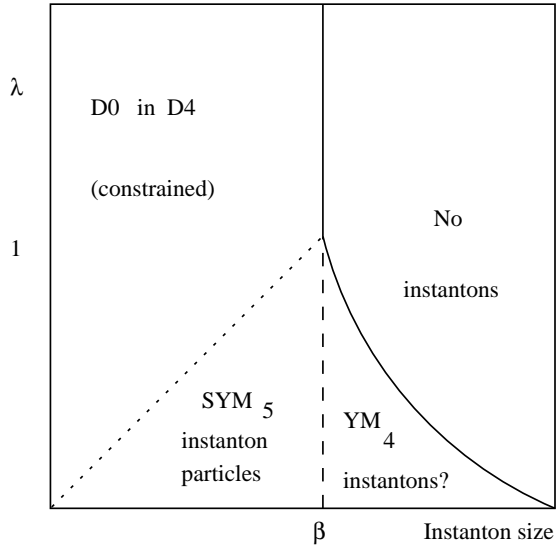}
\caption{
\small
Instanton phase diagram for the compactified D4 theory on a
{\it thermal} circle of size $\beta$. 
  We have
continued the glueball mass scale curve $\rho  \Lambda_{\rm QCD} \sim 1$
to weak coupling in a way
tentatively consistent with asymptotic freedom. 
}
\end{figure}

\section{The Smeared Instanton Solution}
\setcounter{equation}{0}

In the previous section we have seen that probe D0-branes wrapping the
thermal circle of a black D4-brane in the far ultraviolet
 are dual to (unstable)
small-size instantons. Vice-versa, there exists a different 
supergravity solution
dual to a field-theory configuration which can be interpreted 
as containing a condensate of {\it large\/} instantons.

Indeed, the smeared, black D0/D4-brane solution is interpreted 
(as in ref.\ \cite{liutsey} for the supersymmetric T-dual case) 
to be dual to a 
Yang-Mills theory with a non-vanishing
self-dual background. The self-duality of the background implies that
it can be related to instantons, and the smeariness of the D0-branes can 
be interpreted very heuristically as the fact that the instantons are 
`smooth' and then `large'.

 In fact,  in real time,  the
D0-branes are smeared on the D4-branes as soon as they `fall behind' the
horizon, due to the no-hair property (this corresponds $u=u_0$ or, using
 (\ref{sizep}), to $\rho =\beta$.)
This statement has only a heuristic value because, in the euclidean time  
configurations we are considering, space-time effectively ends at
$u=u_0$. Still, the effects of the source D0-branes can be detected
on the long-range fields such as the  metric, dilaton, and RR fields.   
  In this section, we pursue this view of the   smeared  
D0-branes not as probes, as in the previous section, but as background 
data.

The string-frame metric outside a system of $k$ D0-branes smeared over
the volume of  $N$ D4-branes 
differs from that in (\ref{fulmet}) by
one more harmonic function $H_0$:
\be
ds^2 = H_0^{-\half} \,H_4^{-\half} \,h\,d\tau^2 + H_0^{\half}\, H_4^{
-\half}
\,d{\vec y}^{\,2} + H_0^{\half} \,H_4^{\half} \left(dr^2 / h+r^2
\,d\Omega_4^2\right)
.\ee
In the gauge-theory limit, this function is given by 
\be
\label{acheo}   
H_0 (u) = 1+(u_{Q0}/u)^3=
1-\shalf (u_0 /u)^3 +\sqrt{{\mbox{${1\over 4}$}}(u_0 /u)^6 + (u_k
/u)^6}.
\ee
It depends on a new  energy scale $u_k$,
 related to the number density  of D0-branes  $k/V\equiv \kappa$ by
\be
\label{uk}
u_k^3 = \kappa \,{2\pi^3\,\beta \,\lambda  \over N}
.\ee
The new scale is small $(u_k^3
= \ord (1/N))$ in the large-$N$ limit with fixed instanton charge density.
 In this
paper, we are interested in the physics at energies of $\ord (1)$ in the
large-$N$ limit, so that $u_k \ll u_0$ and  $H_0$ may be approximated
by~\footnote{At very low energies, $u_0 \ll u_k$, the smeared solution
is ${\bf X}_6 \times {\bf T}^4$, with ${\bf X}_6$ conformal to
 ${\bf AdS}_2  \times {\bf S}^4$ in the sense of \cite{kostas}.
It is presumably  related to  quantum mechanics in the large-$k$ 
instanton moduli space \cite{seimm}.} 
$
H_0 = 1 +(u_k^2 / u_0 \,u)^3 + \ord (1/N^4).
$
The dilaton profile also receives $\kappa$-dependent corrections,    
$ 
g_s \,e^{-\phi} = (H_4 / H_0^3)^{1/4}
$, 
as well as the Hawking temperature:
\be
\label{hatt}
T^{-1} = \beta= {4 \pi \over 3}\,  
 r_0 \,\sqrt{H_0 (r_0) \, H_4 (r_0)}
=\left({2\pi\lambda\beta\over 9\,u_0} \;H_0 (u_0)\right)^{1/2}.\ee
This yields an equation for $u_0$  that can be
solved iteratively in powers of $(u_k /\lambda T)^6$. 

The relation between  the smeared D0-brane number density $\kappa$ and
the running theta angle is obtained  from the supergravity solution
for the RR two form:
\be
\label{totp}
f_{\rm D0}={c\,\kappa \over u^4} {1\over (H_0)^2} \,du \wedge d\tau
,\ee
with $c$ a known numerical constant. 
As before, a wrapped D0-brane probe can be used to measure an  effective
theta angle whose value at $u=\infty$ defines the bare theta angle.
Plugging (\ref{totp}) into (\ref{thetaf}) we obtain:
\be
\label{thotra}
\theta_{\rm eff} (u) = \int_{{\bf D}_u} f_{\rm D0}=
 \theta \,{u^3 -u_0^3 \over u^3 + u_{Q0}^3}=
\theta \,{h(u)  \over H_0 (u)}, 
\qquad {\rm where}\qquad
\theta = {\beta \,c\,\kappa \over  3 } {1\over u_0^3 + u_{Q0}^3}
.\ee

The two-form solution found by Witten (\ref{witprof}) corresponds to
the  $u_0 \gg u_k$ regime of (\ref{totp}). 
 This provides a relation
between the number density $\kappa$ of 
smeared D0-branes and the bare theta angle,
valid in the large-$N$ limit with fixed $\kappa$:
\be
\label{match}
\theta = {\beta \,c\,\kappa \over  3 \, u_0^3 \,H_0 (u_0)} 
=\kappa\cdot {9\,c\over \lambda^3 \,T^4} \cdot \left({3\over
2\pi}\right)^3 
 + \ord (1/N^2),\ee
where we have used $u_0 =2\pi \lambda T/9  
+ \ord (1/N^2)$,  from (\ref{hatt}). In interpreting this relation, it
is
important to remember that we are working in the $n=0$ vacuum, out of
the $\ord (N)$ metastable vacua mentioned in section 2.2, \ie the
actual values of the parameters are such that the r.h.s of (\ref{match})
is smaller than $2\pi$.  
   
Equation (\ref{match}) is  a very suggestive relation, holographic in nature, 
in which the bare theta angle is obtained in a `mean-field' picture
from the parameters of a kind of `instanton condensate'. We should
stress that (\ref{match}) is only valid in the non-supersymmetric case.
The extremal (supersymmetric) solution has a non-contractible     
${\bf S}^1$ so that we can add an arbitrary harmonic piece to $C_{\rm D0}$,
thereby changing the asymptotic value of $\theta$  independently of
$k$ and $\beta$ (\ie we cannot use Stokes's theorem as we do 
in (\ref{thetaf}) and (\ref{thotra})).

An  interesting application of this connection is the computation of
topological charge correlations to the leading order
in the large-$N$ and large $\lambda$ limit. 
 In view of (\ref{match}), this can be done by
studying the $\kappa$-dependence of the vacuum energy of the
${\rm YM}_4$ theory (or equivalently the thermal free energy of 
the ${\rm SYM}_5$ theory.)  
For example, the action can be calculated as
$ I= \beta \,E_{\rm YM} - S_{\rm BH}$, 
with $S_{\rm BH}=(A_{\rm horizon})/4G_{10}$ the black-brane entropy  and  
$E_{\rm YM} = M_{\rm ADM} - N\,V \,T_{\rm D4}$, the ADM mass above
extremality. One obtains
\be
\label{ac}
I={3\,{\rm Vol} ({\bf S}^4) \,\beta V \over 16\pi G_{10}} \,r_0^3
\left(H_0 (r_0) - {7\over 6}\right) =
N^2 \,{4VT \over \pi \lambda^2} \,u_0^3 \left( H_0 (u_0) - {7\over
6}\right)
.\ee      
Solving $\theta$  from (\ref{match}) and 
using the  relation
\be
\left({u_k \over u_0}\right)^3 = {6\pi^3 \over c} \,H_0 (u_0) \cdot
{\lambda \,\theta \over N}
,\ee
combined with (\ref{acheo}) and (\ref{hatt}), we learn that the
functional
form of the $n=0$ vacuum energy is  given by  
\be
\label{ff}
I(\theta)_{n=0} = N^2 \,VT^4 \, \lambda \,f(\lambda \theta/N)
,\ee
with $f(x)$ an even function (as expected from considerations of
CP symmetry), whose Taylor expansion around $\theta=0$ may be determined
by solving (\ref{hatt}) iteratively. 
These selection rules determine the large-$N$ and large $\lambda$ scaling
of the topological charge correlators at $\theta=0$:        
\be
\left\langle \;(Q_{\rm top})^{2m} \;\right\rangle^{\theta=0}_{\rm
 connected}   \,=\,\left({d \over
 d\theta}\right)_{\theta=0}^{2m}
 \,I(\theta)\,\sim  \,VT^4\; {\lambda^{2m+1}\over N^{2m-2}}  
.\ee
For the standard topological susceptibility, $m=1$, the scaling agrees
 with ref.\ \cite{hashi}.

\section{Concluding Remarks}

 Within the AdS/CFT correspondence, the large-$N$ master field
of the gauge theory is encoded in the gravitational saddle-points
of the supergravity description, subject to boundary conditions.

In the  model of ref. \cite{wit2}, which has  
 a good supergravity description for large  
$N$ and large 't Hooft coupling $\lambda=g^2 N$, 
there are two `master fields', or generalized large-$N$ saddle-points,
given by the two manifolds ${\bf X}_{\rm vac}$ and ${\bf X}_{\rm bb}$, 
with ${\bf S}^1 \times {\bf R}^4$ boundary. We find that ${\bf X}_{\rm
vac}$ supports instantons in the form of wrapped D0-branes, and leads  
to exponentially suppressed theta-angle dependence, very much like
in the ${\bf AdS}_5 \times {\bf S}^5$ case, to which it is dual
through a set of T-duality and localization transitions that we
discuss in some detail, including the matching of the single-instanton
measure.  

However, ${\bf X}_{\rm vac}$ is only the dominant master field in the
supersymmetric case. The
large-$N$ dynamics in the non-supersymmetric case is dominated by
${\bf X}_{\rm bb}$, which {\it does not} support finite-action 
topological
excitations with the instanton charge. Therefore, the dominant master
field shows perturbative (in $1/N$) theta-angle dependence, but has
no `instanton topology', very much like in the two-dimensional toy
models of refs.  
 \cite{toym}. Instead, we can identify approximate (constrained)
instantons of size $\rho \ll \beta$, merging with the vacuum
at sizes of the order of the glueball's Compton wave-length $\rho \sim
\beta$, which for this model coincides with the Kaluza--Klein threshold.

The approximate equivalence of ${\bf X}_{\rm vac}$ and ${\bf X}_{\rm
bb}$ in the ultraviolet regime $u\rightarrow \infty$, poses the question of
whether the approximate small instantons of ${\bf X}_{\rm bb}$ are
really artifacts of the regularization of the Yang--Mills theory
by a hot five-dimensional supersymmetric theory.         
Unfortunately, this question cannot be settled with present techniques,
since $M_{\rm glue} \sim M_{\rm KK}$ in the supergravity
approximation, $\lambda \gg 1$, and we lack a regime in which we
could follow the instantons as genuine  four-dimensional configurations.
 It would be very interesting to see
if the non-supersymmetric gravity
 duals based on Type 0 D-branes \cite{tipocero}
 provide a more
vantageous point to study this question.  

 Heuristically, according to the UV/IR relation, an instanton
of size $\rho \gg \beta$ would be associated to a D0-brane `inside'
the horizon of the black D4-brane. Because of the no-hair properties,
such a configuration would have the D0-charge completely de-localized
over the horizon (see \cite{mpeet} for a recent discussion 
in the extremal case).
 Therefore, such configurations should be interpreted
as homogeneous self-dual backgrounds in the gauge theory, and the
supergravity description involves the `smeared' D0/D4 solution. Although
this picture cannot be held literally in the euclidean solutions, which
lack an `interior region' behind the horizon, we can still identify the
RR two-form generated by  the   
D0-branes `dissolved' in the D4-brane horizon.  
 This RR flux is in turn  responsible for the generation of a 
theta angle, via the AdS/CFT rules of \cite{gkpw}. Therefore, we obtain a
holographic relation between the theta angle
and the smeared instanton charge. Although the general relation between
background fields and theta angle is not new (see \cite{cole, toym} for 
explicit two-dimensional examples), we find it interesting that in our
case the background field is explicitly associated to an instanton
condensate, with quantized topological charge (equal to the number $k$ of
smeared D0-branes). This is 
reminiscent of the instanton liquid models, where
the instanton density is fixed self-consistently (see for instance
\cite{shu}).

\section*{Acknowledgements}

We would like to thank Margarita Garc\'{\i}a P\'erez, 
Yaron Oz and Kostas Skenderis for useful discussions. 
This work is partially supported by the European Commission TMR
programme ERBFMRX-CT96-0045 in which J.L.F.B.\ is associated to the
University of Utrecht and A.P.\ is associated to the Physics Department,
University of Milano. 
A.P.\ would like to thank CERN for its hospitality while part of this
work was carried out. 

%


\begin{thebibliography}{99}
{\small 
\bibitem{shif} M. Shifman, Ed. {\it `Instantons in Gauge Theories'}, 
World Scientific (1994). 
 
\bibitem{tof} G. 't Hooft, \npb{72}{1974}{461.}

\bibitem{venewitt} E. Witten, \npb{156}{1979}{269;} {\it Ann. 
Phys.} {\bf 128} (1980) 363. 
G. Veneziano, \npb{159}{1979}{213.}


\bibitem{toym} A. D'Adda, M. Luscher and P. DiVecchia,
\npb{146}{1979}{63.}
E. Witten, \npb{149}{1979}{285.}

\bibitem{Malda} J. Maldacena,
\atmp{2} {1998} {231,}
\bb{9711200.}

\bibitem{gkpw} S. Gubser, I. Klebanov and A. Polyakov,
\plb{428} {1998} {105,}
\bb{9802109.}
 E. Witten,
\atmp{2} {1998} {253,} \bb{9802150.}
 
\bibitem{dinst} T. Banks and M.B. Green, \jhep{9805}{1998}{002,}      
\bb{9804170.} C-S. Chu, P-M. Ho and Y-Y. Wu, \npb{541}{1999}{179,}
\bb{9806103.} I.I. Kogan and G. Luz\'on, \npb{539}{1999}{9806197.} M.
Bianchi, M.B. Green, S. Kovacs and G. Rossi, \jhep{9808}{1998}{013,}
\bb{9807033.}


\bibitem{wit2} E.\ Witten,  \atmp{2}{1998}{505,} \bb{9803131.}
 
\bibitem{usinst} J.L.F. Barb\'on and  A. Pasquinucci,
\npb{517} {1998} {125,} \bb{9708041;}
{\it Fortsch. Phys.} {\bf 47} (1999) 255, \bb{9712135.}

\bibitem{mali} N. Itzhaki, J. Maldacena, J. Sonnenschein and S.
Yankielowicz, \prd{58}{1998}{046004,} \bb{9802042.}

\bibitem{krev} K. Skenderis, \bb{9901050.}

\bibitem{bkr} J.L.F. Barb\'on, I.I. Kogan and  E. Rabinovici,  
\npb{544} 
{1999} {104,} \bb{9809033.} 

\bibitem{hp} S.W. Hawking and  D. Page,
\cmp{87} {1983} {577.}
 
\bibitem{oo} C. Csaki, H. Ooguri, Y. Oz and J. Terning,
\jhep{9901}{1999}{017,} \bb{9806021.} 

\bibitem{stten} A. Brandhuber, N. Itzhaki, J. Sonnenschein and S.
Yankielowicz, \jhep{9806}{1998}{001,} \bb{9803263.} D.J. Gross and H.
Ooguri, \prd{58}{1998}{106002,} \bb{9805129.} 

\bibitem{koo} H. Ooguri and K. Skenderis, \jhep{9811}{1998}{013,}  
\bb{9810128.} 
     
\bibitem{uvir}  L. Susskind and  E. Witten,
\bb{9805114.}
A. Peet and J. Polchinski, 
\prd{59} {1999} {065006,} \bb{9809022.} 

\bibitem{greglaf} R. Gregory and  R. Laflamme,
\prl{70} {1993} {2837,} \bb{9301052.}

\bibitem{horpol} G.T. Horowitz and J. Polchinski,  
\prd{55} {1997} {6189,} 
\bb{9612146.} 

\bibitem{valya} N. Dorey, T. Hollowood, V.V. Khoze, M. Mattis and S.
Vandoren, \bb{9901128.} 

\bibitem{thetaw} E.\ Witten, 
\prl{81} {1998} {2862,}  
\bb{9807109.} 


\bibitem{ozpa} J.L.F. Barb\'on and A. Pasquinucci, \plb{421}{1998}{131,}
\bb{9711030.}  
 Y. Oz and  A. Pasquinucci,  
\plb{444} {1998} {318,}  \bb{9809173.}  


\bibitem{liutsey} H. Liu and  A.A. Tseytlin, 
\bb{9903091.} 

\bibitem{kostas} H.J. Boonstra, K. Skenderis and P.K. Townsend, 
\jhep{9901}{1999}{003,}  
\bb{9807137.}  
\bibitem{seimm} O. Aharony, M. Berkooz, S. Kachru, N. Seiberg and E.
Silverstein, \atmp{1}{1998}{148,} \bb{9707079.} 

\bibitem{hashi} A. Hashimoto and Y. Oz, \bb{9809106}

\bibitem{tipocero} I Klebanov and A.A. Tseytlin, \bb{9811035.} 

\bibitem{mpeet} D. Marolf and A. Peet, \bb{9903213.} 

\bibitem{cole} S. Coleman, {\it Ann.  Phys.} {\bf 101} (1976) 239.

\bibitem{shu} T. Schafer and E.V. Shuryak, {\it Rev. Mod. Phys.} {\bf 70}
(1998) 323,
\bb{9610451.}   
} 
\end{thebibliography}
\end{document}